\documentclass[12pt,letter]{article}
\pdfoutput=1
\usepackage{graphicx, epsfig, color,cite}
\usepackage{amsmath}
\usepackage{amssymb}
\usepackage{caption,subcaption,graphicx}
\usepackage[hidelinks]{hyperref}
\def\to{\rightarrow}

\def\bi{\begin{itemize}}
\def\ei{\end{itemize}}

\def\tf{\tilde f}

\def\tst{\tilde t}

\def\tg{\tilde g}

\def\tell{\tilde\ell}
\def\tq{\tilde q}

\def\tw{\widetilde\chi^{\pm}}
\def\twns{\widetilde\chi}
\def\tz{\widetilde\chi^0}
\def\alt{\lesssim}
\def\agt{\gtrsim}
\def\be{\begin{equation}}  
\def\ee{\end{equation}}  
\def\bea{\begin{eqnarray}}  
\def\eea{\end{eqnarray}}

\begin{document}
\begin{titlepage}
\begin{flushright}
OU-HEP-200730
\end{flushright}

\vspace{0.5cm}
\begin{center}
{\Large \bf The LHC higgsino discovery plane\\
for present and future SUSY searches
}\\ 
\vspace{1.2cm} \renewcommand{\thefootnote}{\fnsymbol{footnote}}
{\large Howard Baer$^1$\footnote[1]{Email: baer@ou.edu },
Vernon Barger$^2$\footnote[2]{Email: barger@pheno.wisc.edu},
Shadman Salam$^1$\footnote[3]{Email: shadman.salam@ou.edu}\\
Dibyashree Sengupta$^1$\footnote[4]{Email: Dibyashree.Sengupta-1@ou.edu} 
and Xerxes Tata$^3$\footnote[5]{Email: tata@phys.hawaii.edu}
}\\ 
\vspace{1.2cm} \renewcommand{\thefootnote}{\arabic{footnote}}
{\it 
$^1$Homer L. Dodge Department of Physics and Astronomy,\\
University of Oklahoma, Norman, OK 73019, USA \\[3pt]
}
{\it 
$^2$Department of Physics,
University of Wisconsin, Madison, WI 53706 USA \\[3pt]
}
{\it 
$^3$Department of Physics and Astronomy,
University of Hawaii, Honolulu, HI, USA\\[3pt]
}

\end{center}

\vspace{0.5cm}
\begin{abstract}
\noindent
Considerations from electroweak naturalness and stringy naturalness
imply a little hierarchy in supersymmetric models where the
superpotential higgsino mass parameter $\mu$ is of order the weak scale
whilst the soft SUSY breaking terms may be in the (multi-) TeV range.
In such a case, discovery of SUSY at LHC may be most likely in the
higgsino pair production channel.  Indeed, ATLAS and CMS are performing
searches in the higgsino mass discovery plane of $m_{\tz_2}$ vs. $\Delta
m^0\equiv m_{\tz_2}-m_{\tz_1}$.  We examine several theoretical aspects
of this discovery plane in both the gravity-mediation NUHM2 model and
the general mirage-mediation (GMM$^\prime$) models.  These include: the
associated chargino mass $m_{\tw_1}$, the expected regions of the
bottom-up notion of electroweak naturalness $\Delta_{EW}$, and the
expected regions of stringy naturalness. 
While compatibility with electroweak naturalness allows for mass gaps 
$\Delta m^0\sim$ 4-20~GeV, 
stringy naturalness exhibits a clear preference for 
yet smaller mass gaps of 4-10~GeV.
For still smaller mass gaps, the plane becomes sharply unnatural
since very large gaugino masses are required.  This study informs the
most promising SUSY search channels and parameter space regions for the
upcoming HL-LHC runs and possible HE-LHC option.

\end{abstract}
\end{titlepage}


The discovery of the Higgs boson with mass $m_h\simeq 125$ GeV at the
LHC\cite{atlas_h,cms_h} is enigmatic within the context of the Standard
Model (SM) in that $m_h$ exhibits quadratic sensitivity to the highest
mass scale (such as the Grand Unification scale) that the SM might couple
to: radiative corrections would then
drive its mass to far higher values.  
The introduction of softly broken
supersymmetry (SUSY) 
results in cancellations leaving only  logarithmic sensitivity
to the scale of new physics, 
and the Higgs mass can be
stabilized at its measured value\cite{witten_kaul}.  Weak scale
SUSY\cite{wss} finds a natural home within string theory, and is the
oft-sought weak scale realization of string
compactifications.\footnote{For recent discussion, see {\it e.g.}
  Ref. \cite{Broeckel:2020fdz}.}  Weak scale SUSY is actually supported by a
variety of {\it virtual} quantum effects, including 1. the measured
values of the gauge couplings are consistent with unification within the
context of the minimal supersymmetric standard model (MSSM)\cite{gauge},
2. the measured value of the top mass is just right to radiatively drive
electroweak symmetry breaking in the MSSM\cite{rewsb}, 3. the measured
value of $m_h$ lies within the narrow window of predicted values in the
MSSM\cite{mhiggs} and 4. precision EW observables, especially within the
$m_W$ vs. $m_t$ plane, slightly favor heavy SUSY over even the
SM\cite{sven}.

Even so, searches for SUSY at LHC Run 2 with 139 fb$^{-1}$ of integrated
luminosity\cite{Canepa:2019hph} have led to limits (within the
context of various simplified models) of $m_{\tg}\agt 2.2$
TeV\cite{lhc_gl} and $m_{\tst_1}\agt 1.1$ TeV\cite{lhc_t1}.  In this
case, one expects the corresponding soft SUSY breaking terms to lie in
the (multi) TeV range.  But if the soft SUSY breaking parameters are too
large, then a Little Hierarchy (LH) emerges: one might expect the Higgs
mass to be of order the soft breaking scale.  This is exemplified by the
fact that the LHC SUSY particle mass limits lie far beyond initial
estimates from naturalness wherein values such as
$m_{\tg},\ m_{\tst_1}\alt 400$ GeV were expected\cite{eenz,bg,dg,ac}.
In retrospect, it was pointed out that the log-derivative measure
$\Delta_{BG}\equiv max_i |\frac{\partial\log m_Z^2}{\partial\log p_i}|$,
where the $p_i$ are fundamental parameters (usually the soft breaking
terms are taken as the $p_i$) of the 4-d low energy effective SUSY
theory, is highly model-dependent\cite{DEW,mt,seige,arno}.  In a
top-down approach within a more UV complete theory, such as string
theory, then all soft terms are (in principle) calculable in terms of
more fundamental parameters (such as the gravitino mass $m_{3/2}$ in
gravity or anomaly--mediation), and the value of $\Delta_{BG}$ changes
greatly from its effective theory value\cite{arno,midi}.  Alternatively,
in the string theory landscape-- wherein the soft terms may scan within
the landscape-- then selection effects may operate so that certain
ranges of soft term values are statistically preferable to
others\cite{land1,land2,land3,land4}.

A {\it model independent} bottom-up
measure of electroweak naturalness emerges directly from
minimizing the scalar potential of the MSSM in order to relate the Higgs field 
vevs to the MSSM Lagrangian parameters. 
The electroweak fine-tuning parameter~\cite{ltr,rns}, $\Delta_{\rm EW}$,
is a measure of the degree of cancellation between various contributions
on the right-hand-side (RHS) in the well-known expression for the $Z$ mass:
\be \frac{m_Z^2}{2} = \frac{m_{H_d}^2 +\Sigma_d^d -(m_{H_u}^2+\Sigma_u^u)\tan^2\beta}
{\tan^2\beta -1} -\mu^2\simeq  -m_{H_u}^2-\Sigma_u^u-\mu^2
\label{eq:mzs}
\ee
which results from the minimization of the Higgs potential in the MSSM.
Here, $\tan\beta =v_u/v_d$ is the ratio of Higgs field
vacuum-expectation-values and the $\Sigma_u^u$ and $\Sigma_d^d$
contain an assortment of radiative corrections, the largest of which
typically arise from the top squarks. Expressions for the $\Sigma_u^u$
and $\Sigma_d^d$ are given in the Appendix of Ref.~\cite{rns} and are included
in the Isajet SUSY spectrum generator\cite{isajet}.
We also include leading two-loop terms from $m_{\tg}$ and $m_{\tst_{1,2}}$ as 
determined by Dedes and Slavich\cite{Dedes:2002dy}.
If the RHS terms in Eq.~(\ref{eq:mzs}) are individually
comparable to $m_Z^2/2$, then no unnatural fine-tunings are required to
generate $m_Z=91.2$ GeV. $\Delta_{\rm EW}$ is defined to be the largest
of these terms, scaled by $m_Z^2/2$. Clearly, low electroweak
fine-tuning requires that $\mu$ be close to $m_Z$ and that $m_{H_u}^2$--
which sets the values of $m_{W,Z,h}$--
be radiatively driven to {\it small} negative values close to the weak scale.
This scenario has been dubbed radiatively-driven natural supersymmetry
or RNS~\cite{ltr,rns} since it allows for large, seemingly unnatural GUT scale
soft terms to be driven to natural values at the weak scale via RG running.

An advantage of $\Delta_{EW}$ is its model independence in that it
depends only on weak scale Lagrangian parameters and sparticle
masses. Thus, for a given mass spectrum, one obtains the same value of
$\Delta_{EW}$ whether it was generated in some high scale model such as
mSUGRA or else just within the pMSSM: {\it i.e.} it is both parameter
independent and scale independent.  If one moves to models with extra
low-scale exotic matter beyond the MSSM, then additional terms may have
to be added to the RHS of Eq.~(\ref{eq:mzs}). It has been argued that
$\Delta_{BG} \rightarrow \Delta_{EW}$ if appropriate underlying
correlations between model parameters-- 
that are usually assumed to be independent-- 
are incorporated \cite{DEW,mt}.

Under $\Delta_{EW}$, the natural SUSY parameter space is found to be far
larger than what is expected under $\Delta_{BG}$\cite{land4}.  Since
top-squarks enter Eq.~(\ref{eq:mzs}) at one-loop level, they can have
masses into the several TeV regime while remaining natural, with
$\Delta_{EW}\alt 30$.\footnote{The onset of large finetuning for values
  of $\Delta_{EW}>30$ is visually displayed in Fig. 1 of
  Ref. \cite{upper}.}  Gluinos, which enter Eq.~(\ref{eq:mzs}) at two-loop
level\cite{Dedes:2002dy}, can range up to $\sim 6$ TeV at little cost to
naturalness\cite{upper,gainer27}.  But since the SUSY conserving $\mu$
parameter enters Eq.~(\ref{eq:mzs}) directly, then the lightest Higgs
boson and the superpartner higgsinos must have mass not too far removed
from $m_{weak}\sim m_{W,Z,h}\sim 100$ GeV.  Thus, we expect from
$\Delta_{EW}$ that higgsinos will be the lightest superpartners while
other sparticles which gain mass from soft breaking terms may well be
beyond the present LHC mass limits.

Such a scenario, it has been suggested\cite{land1,land2}, 
emerges naturally from the landscape of string theory vacua 
which also provides a solution to the cosmological constant problem. 
Rather general considerations of the string theory landscape lead to an expected
distribution of soft terms for different pocket-universes within the
multiverse which favors large values by a power law distribution:
$m_{soft}^{2n_F+n_D-1}$ where $n_F$ is the number of $F$-term fields and
$n_D$ is the number of $D$-term fields contributing to the overall SUSY
breaking scale\cite{Douglas:2004qg}. 
However, the overall SUSY breaking scale $m_{soft}$
cannot be too large lest it lead to too large a value of $m_{weak}^{PU}$
in different pocket universes (PU).  The atomic
principle\cite{Donoghue:2007zz} -- that atoms as we know them ought to
exist in a pocket-universe which gives rise to observers -- requires
that $m_{weak}^{PU}$ be within a factor 2-5 of the measured value of
$m_{weak}^{OU}$ in our universe (OU).  If the value of $\mu$ is
determined by whatever solution to the SUSY $\mu$ problem is
invoked\cite{Bae:2019dgg}, then $\mu$ is {\it unavailable} for (the
usual) electroweak fine-tuning and the value of $m_{weak}^{PU}$ is
determined by Eq.~(\ref{eq:mzs}).  The requirement that $m_Z^{PU}\alt
4m_Z^{OU}$ is then equivalent to the above mentioned naturalness
requirement that $\Delta_{EW}<30$.  Thus, the concept of {\it stringy
  naturalness}\cite{douglas,land4} favors soft terms as large as
possible such that the weak scale remains not too far from its measured
value in our universe.  Under such conditions, superpartners are lifted
beyond LHC search limits while the light Higgs mass $m_h$ is pulled to a
statistical peak at $\sim 125$ GeV\cite{land1,land2}.  In particular,
the gluino mass is expected at $m_{\tg}\sim 4\pm 2$ TeV while
$m_{\tst_1}\sim 1.7\pm 1$ TeV\cite{land2}.  First/second generation
matter scalars are pulled into the $m_{\tq,\tell}\sim 25\pm 15$ TeV
range leading to a mixed decoupling/quasi-degeneracy solution to the
SUSY flavor and CP problems\cite{Baer:2019zfl}.  Under such (highly
motivated) conditions, most sparticles may well lie beyond the reach of
high-luminosity LHC (HL-LHC), with $\sqrt{s}\sim 14$ TeV and integrated
luminosity $\sim 3$ ab$^{-1}$.  The exception is the four higgsinos
$\tz_{1,2}$ and $\tw_1$ which are expected to have mass $\sim 100-350$
GeV.

The search for higgsino pair production at the LHC is fraught with some
difficulties.  The lightest neutralino $\tz_1$ is expected to form
typically 10-20\% of dark matter\cite{Baer:2013vpa, Baer:2017cck} with
the remainder perhaps being composed of axions\cite{Bae:2013bva}.  Indeed, such
a scenario naturally emerges from the hybrid CCK/SPM
solutions\cite{cck,spm} to the SUSY $\mu$ problem\cite{Baer:2018avn},
where both $R$-parity and the global $U(1)_{PQ}$ symmetry (needed for an
axionic solution to the strong CP problem) emerge as accidental
approximate remnant symmetries from a more fundamental ${\bf Z}_{24}^R$
symmetry (which itself is expected to emerge from compactification of a
10-d Lorentz string symmetry down to 4-d, $N=1$ SUSY effective 
theory\cite{Nilles:2017heg}).
The small mass gaps $\Delta m^0= m_{\tz_2}-m_{\tz_1}$ and $\Delta
m^+\equiv m_{\twns_1^+}-m_{\tz_1}$ (following Guidice \& Pomarol
notation, Ref. \cite{gp}) between the various higgsinos means that
production of $\tz_1\tz_2$, $\tw_1\tz_2$ and $\twns_1^+\twns_1^-$ leads
to very soft visible decay products, and where most of the energy goes
into making the two lightest SUSY particles' (LSP) rest mass.  In
addition, $\tz_1\tz_1 j$ production provides a monojet at the level of
1/100 signal/background, where the dominant background comes from $Zj$
production\cite{Baer:2014cua}.  The reaction $pp\to\tz_1\tz_2$ with
$\tz_2\to \mu^+\mu^-\tz_1$ was proposed in Ref. \cite{bbh} which would
require a soft dimuon trigger to record the events.  In Ref's
\cite{than,kribs}, it was proposed to look at $\tz_1\tz_2j$ production
where an ISR jet radiation at high $p_T\agt 100$ GeV could provide
either a jet or MET trigger. Indeed, ATLAS\cite{Aad:2019qnd} and
CMS\cite{CMS:2016zvj} have followed up on the opposite-sign dilepton
plus jet(s) plus MET signature (OSDJMET), and have provided limits on
such reactions in the $m_{\tz_2}$ vs. $\Delta m^0$ plane.  Due to its
promising prospects for SUSY discovery, we will henceforth label this as
the {\it LHC higgsino discovery plane}.  Indeed, the latest ATLAS
analysis from LHC Run 2 with 139 fb$^{-1}$ finds some {\it excess of
  events} with low dilepton invariant mass $m(\ell^+\ell^-)\sim 5-10$
GeV in their SR-E-med analysis (see Fig. 11{\it a} of Ref. \cite{Aad:2019qnd}).
It will be exciting to see if this excess is confirmed in the
forthcoming CMS 139~fb$^{-1}$analysis, or in future data from LHC Run 3
or HL runs.

The ATLAS and CMS searches within the higgsino discovery plane take
place within simplified models which are appropriate for the OSDJMET
search.  Our goal in this paper is to place the higgsino discovery plane
within the context of natural SUSY models and landscape SUSY models so
as to provide theoretical context for the discovery plane.  For
instance, what features of the plane are model-dependent or
model-independent, and which portions of the plane are favored by
naturalness and by the string theory landscape? Identifying such regions
should help focus OSDJMET searches onto the most promising portions of
parameter space, and also help to prioritize searches in promising
regions over searches within regions with implausible parameter choices.

To compare the simplified model of the higgsino discovery plane against expectations from
theory, we work with two well-motivated models. 
The first is generic supergravity GUTs as portrayed in the two-extra-parameter
non-universal Higgs model (NUHM2)\cite{nuhm2}. 
This model takes similar parameters as the well-known
mSUGRA/CMSSM model except that the two Higgs doublets acquire independent soft terms
$m_{H_u}^2$ and $m_{H_d}^2$ whereas the three generations of matter scalars unify to $m_0$. 
This model is better motivated than mSUGRA/CMSSM since the Higgs multiplets 
necessarily live in different GUT multiplets from matter scalars, 
while the latter may unify in $SO(10)$ SUSY GUTs\cite{raby} 
or in stringy local GUTs\cite{localguts}. 
In NUHM2, the gauginos still unify to $m_{1/2}$ at the GUT scale whilst trilinear 
soft terms unify to $A_0$. For convenience, the GUT values of $m_{H_u}^2$ and $m_{H_d}^2$ 
are traded for weak scale parameters $\mu$ and $m_A$. 
As usual, $\tan\beta$ is the ratio of Higgs vevs. 
Thus, the parameter space is given by
\be
m_0,\ m_{1/2},\ A_0,\ \tan\beta,\ \mu,\ m_A\ \ \ \ (NUHM2) .
\ee
It is easy to generalize this to the NUHM3 or NUHM4 
models where the third generation or each generation separately
acquires an independent soft mass $m_0(i)$.  But for illustration, we will
take the generations as degenerate.\footnote{Within the string theory
  landscape, first/second generation matter scalar masses are pulled to
  a (generation independent) upper bound in the $20\pm 10$ TeV regime,
  offering a mixed decoupling/quasi-degeneracy solution to the SUSY
  flavor and CP problems\cite{Baer:2019zfl}.}

A well-motivated alternative is the generalized mirage-mediation
model\cite{GMM} (GMM) which contains comparable moduli-mediated and
anomaly-mediated contributions to soft terms.  Their relative
contributions are parametrized by $\alpha: 0\to\infty$ where $\alpha\to
0$ gives the pure AMSB\cite{amsb} soft terms and $\alpha\to \infty$
gives pure moduli (gravity) mediation.  It is called mirage mediation
because the gaugino mass universality is offset by AMSB contributions
proportional to the corresponding gauge group beta functions.  Then,
evolution of gaugino masses from $m_{GUT}$ to $m_{weak}$ results in
gaugino mass unification at the mirage
scale\cite{Choi:2005uz,Falkowski:2005ck} \be \mu_{mir}=m_{GUT}\cdot
e^{(-8\pi^2/\alpha )} .  \ee

\begin{figure}[!htbp]
\begin{center}
\includegraphics[height=0.4\textheight]{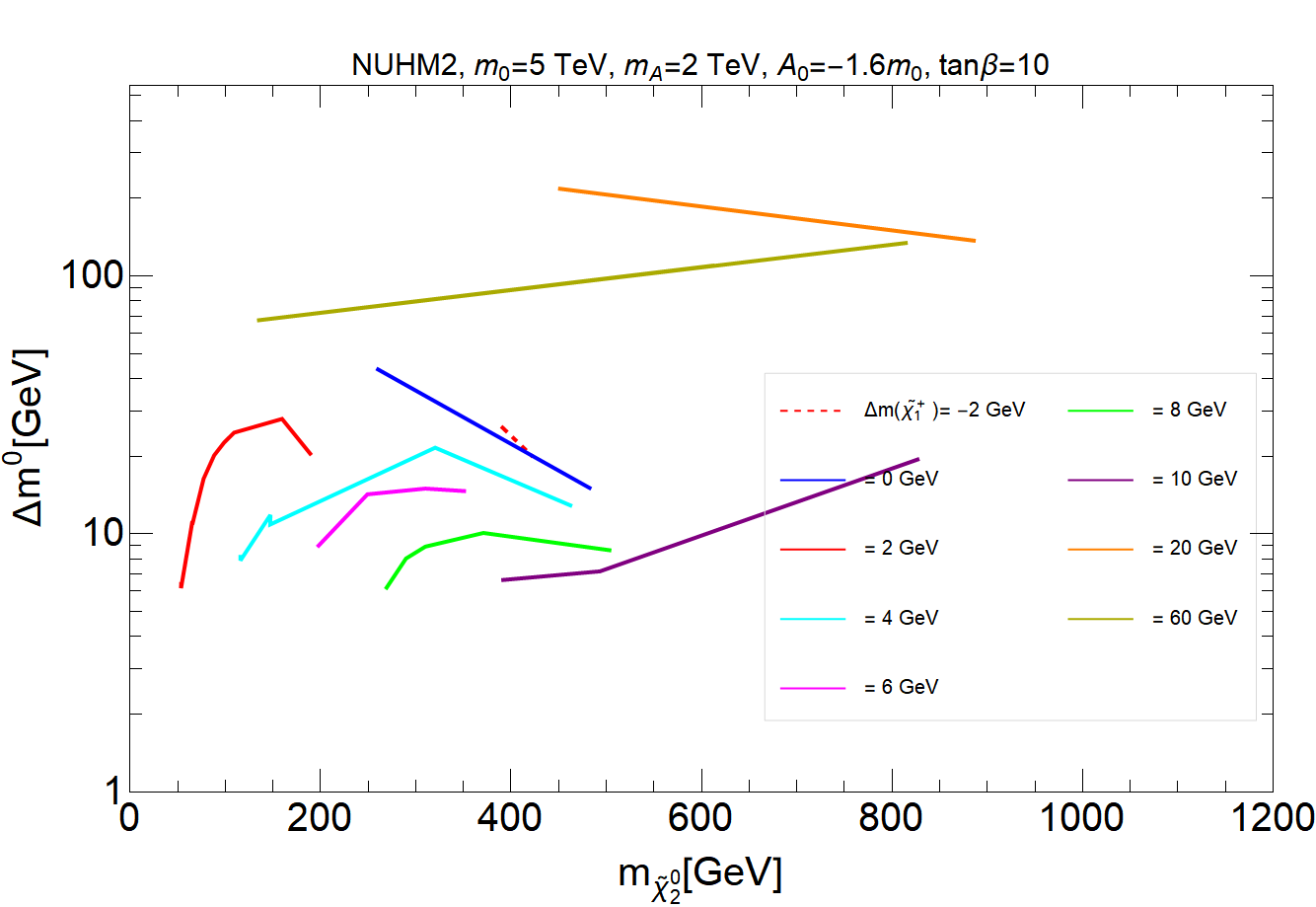}
\includegraphics[height=0.4\textheight]{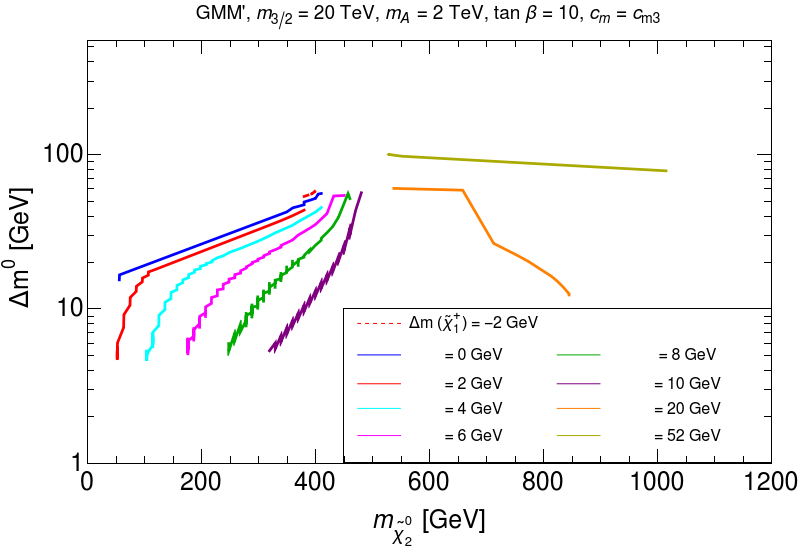}
\caption{Deviations in loop-corrected chargino mass as compared
to simplified model value $\Delta m(\tw_1)$
{\it a}) in the NUHM2 model with varying $\mu$ and $m_{1/2}$ 
but with $m_0=5$ TeV, $A_0=-1.6 m_0$, $\tan\beta =10$ and $m_A=2$ TeV,
and {\it b}) in the GMM$^\prime$ model with varying $\mu$ and 
$\alpha$ but with $m_{3/2}=20$ TeV and $c_m=c_{m3}$.
Both models take $m_A=2$ TeV and $\tan\beta =10$. 
\label{fig:mw1}}
\end{center}
\end{figure}

\begin{figure}[!htbp]
\begin{center}
\includegraphics[height=0.4\textheight]{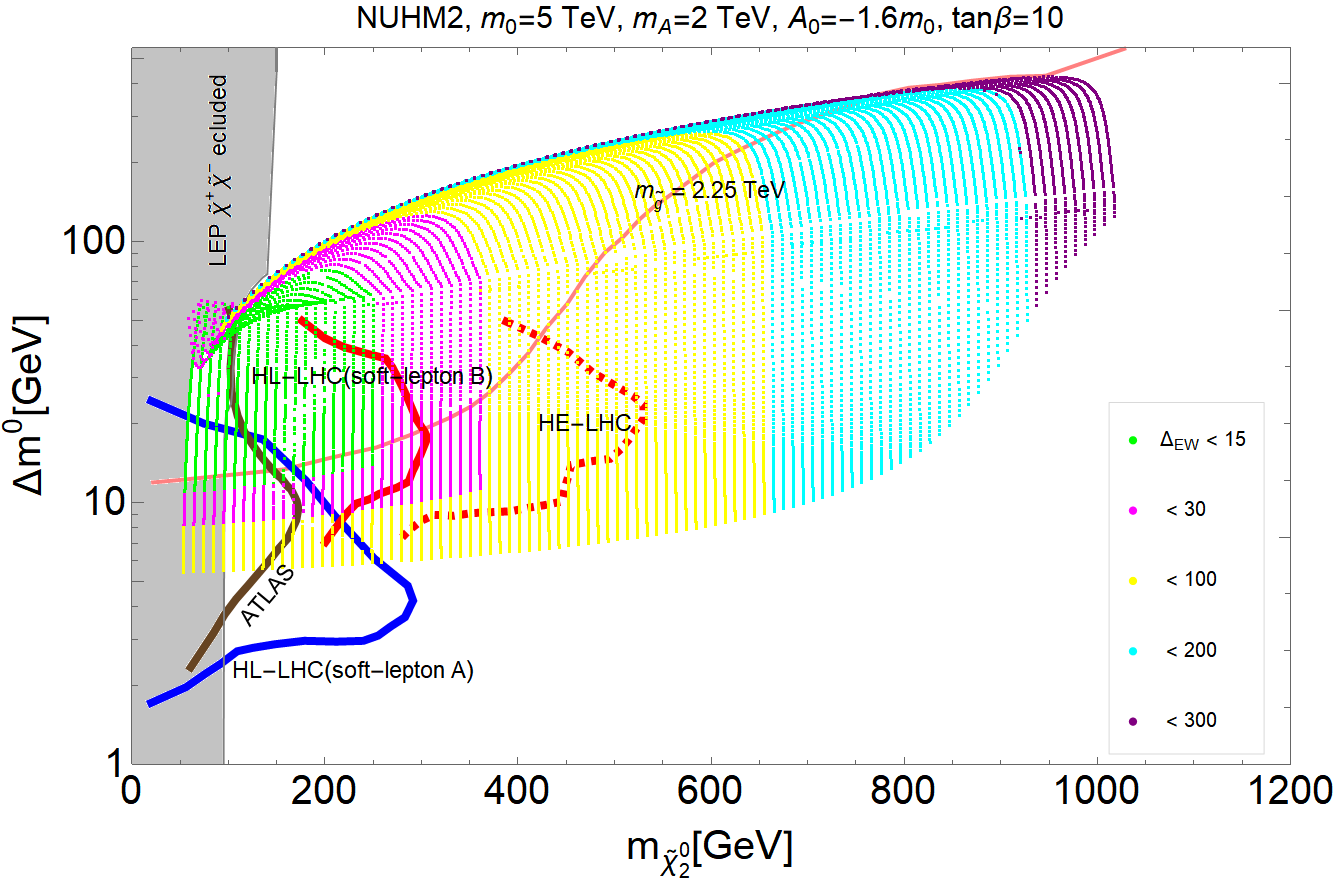}
\includegraphics[height=0.4\textheight]{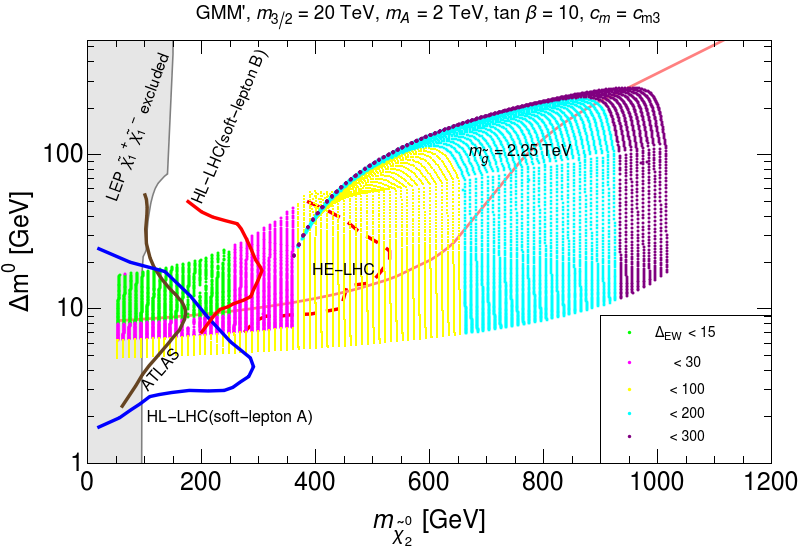}
\caption{Regions of naturalness $\Delta_{EW}$ in the higgsino discovery plane
$m_{\tz_2}$ vs. $\Delta m^0\equiv m_{\tz_2}-m_{\tz_1}$ from {\it a}) the NUHM2 model
with varying $\mu$ and $m_{1/2}$ but with $m_0=5$ TeV, $A_0=-1.6 m_0$ and
{\it b}) the GMM$^\prime$ model with varying $\mu$ and $\alpha$ but with 
$m_{3/2}=20$ TeV and $c_m=c_{m3}$. 
For both models, we take $\tan\beta =10$ and $m_A=2$ TeV. 
We also show the present reach of the ATLAS
experiment with 139 fb$^{-1}$ and the ATLAS (soft lepton A) and CMS (soft lepton B) 
projected future reach at HL-LHC and also CMS at HE-LHC. 
The region above the $m_{\tg}=2.25$ TeV contour is excluded by current LHC Run 2
gluino search analysis. 
The Higgs mass $m_h\sim 125$ GeV throughout the plane while $m_{\tst_1}>1.1$ TeV everywhere.
\label{fig:dew}}
\end{center}
\end{figure}

\begin{figure}[!htbp]
\begin{center}
\includegraphics[height=0.4\textheight]{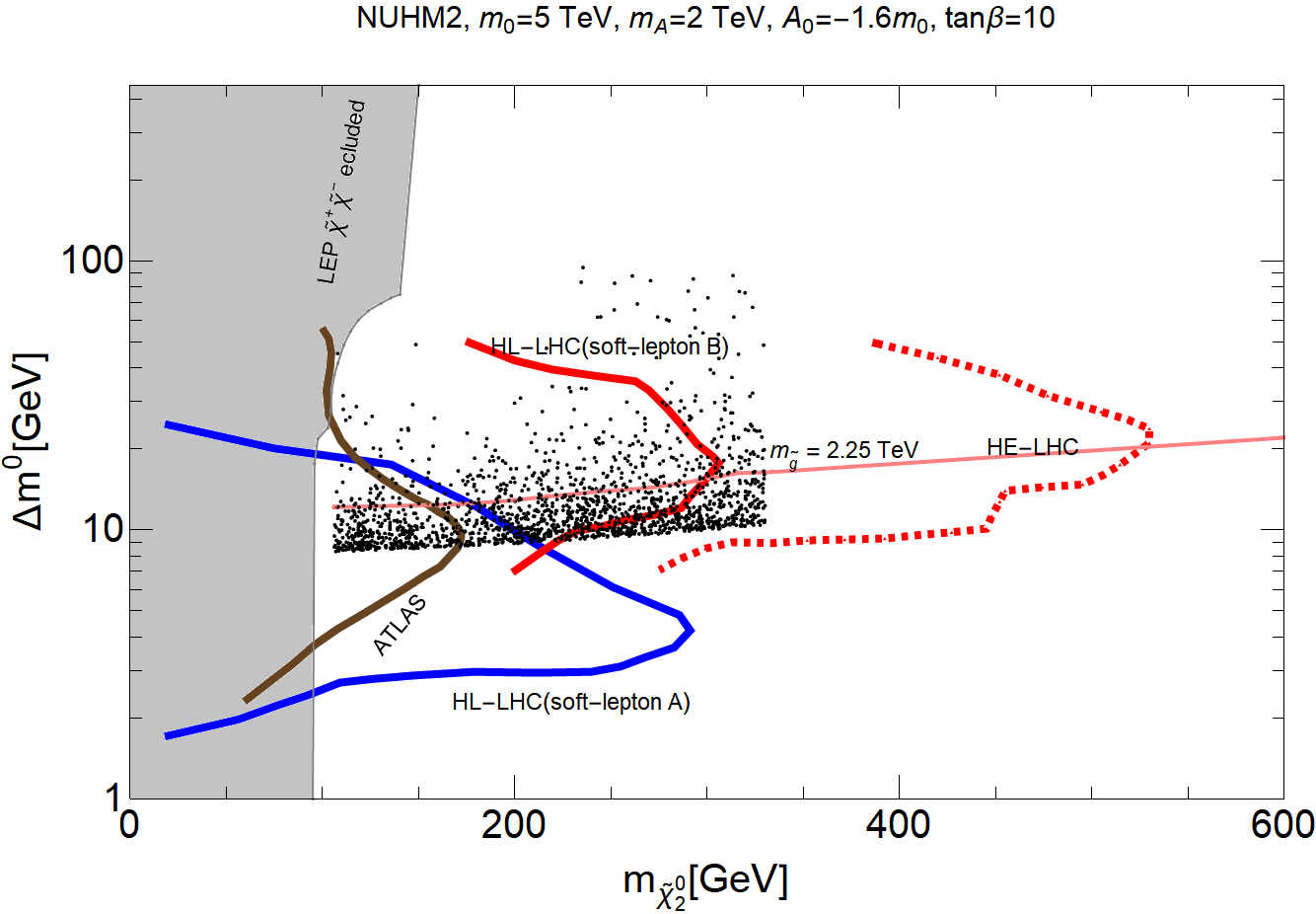}
\includegraphics[height=0.4\textheight]{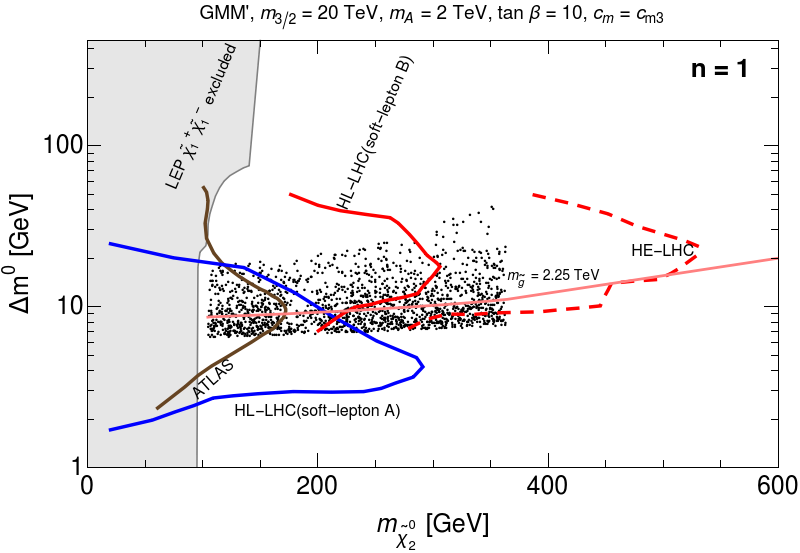}
\caption{Regions of stringy naturalness in the higgsino discovery plane
  $m_{\tz_2}$ vs. $\Delta m^0\equiv m_{\tz_2}-m_{\tz_1}$ for {\it a})
  the NUHM2 model with varying $\mu$ and $m_{1/2}$ but with $A_0=-1.6
  m_0$ and {\it b}) in the GMM$^\prime$ model with varying $\mu$ and
  $\alpha$ but with $m_{3/2}=20$ TeV and $c_m=c_{m3}$.  For both frames,
  we take $\tan\beta =10$ and $m_A=2$ TeV.  We also show the present
  reach of the ATLAS experiment with 139 fb$^{-1}$ and the ATLAS (soft
  lepton A) and CMS (soft lepton B) projected future reach at HL-LHC and
  CMS projected reach at HE-LHC.  The region above the $m_{\tg}=2.25$
  TeV contour is excluded by current LHC Run2 gluino pair searches.  The
  Higgs mass $m_h\sim 125$ GeV throughout the plane while
  $m_{\tst_1}>1.1$ TeV everywhere.
\label{fig:lSUSY}}
\end{center}
\end{figure}

The soft SUSY breaking terms in GMM are given by
\begin{eqnarray}
M_a&=& M_s\left( \alpha +b_a g_a^2\right),\label{eq:M}\\
A_{ijk}&=& M_s \left( -a_{ijk}\alpha +\gamma_i +\gamma_j +\gamma_k\right),
\label{eq:A}\\
m_i^2 &=& M_s^2\left( c_i\alpha^2 +4\alpha \xi_i -
\dot{\gamma}_i\right) ,\label{eq:m2}
\end{eqnarray}
where $M_s\equiv\frac{m_{3/2}}{16\pi^2}$,
$b_a$ are the gauge $\beta$ function coefficients for gauge group $a$ and 
$g_a$ are the corresponding gauge couplings. The coefficients that
appear in (\ref{eq:M})--(\ref{eq:m2}) 
originally appeared as discrete quantities for particular orbifold
compactifications where the $n_i$ are modular weights.
They are given by
$c_i =1-n_i$, $a_{ijk}=3-n_i-n_j-n_k$ and
$\xi_i=\sum_{j,k}a_{ijk}{y_{ijk}^2 \over 4} - \sum_a l_a g_a^2
C_2^a(f_i).$ 
These coefficients are generalized in GMM to adopt continuous values
to allow 
for more generic ways of moduli stabilization and potential
uplifting \cite{GMM}. 
The gaugino mass relations (\ref{eq:M}) are, however, 
much more robust \cite{Choi:2007ka}.
Finally, $y_{ijk}$ are the superpotential Yukawa couplings,
$C_2^a$ is the quadratic Casimir for the a$^{th}$ gauge group
corresponding to the representation to which the sfermion $\tf_i$ belongs,
$\gamma_i$ is the anomalous dimension and
$\dot{\gamma}_i =8\pi^2\frac{\partial\gamma_i}{\partial \log\mu}$.
Expressions for the last two quantities involving the 
anomalous dimensions can be found in the Appendices of 
Ref's.~\cite{Falkowski:2005ck,Choi:2006xb}.
In the GMM model, the coefficients $c_{H_u}$ and $c_{H_d}$ can be traded
for more convenient weak scale values $\mu$ and $m_A$ as in the NUHM2 
model, yielding the GMM$^\prime$ model\cite{GMM} 
with a parameter space given by 
\be
\alpha,\ m_{3/2},\ c_m,\ c_{m3},\ a_3,\ \tan\beta ,\ \mu ,\ m_A
\ \ \ (GMM^\prime ). \label{eq:gmmp}  
\ee
Here, $m_{3/2}$ is the gravitino mass while $c_m$ and $c_{m3}$ vary the 
moduli-to-AMSB contributions for first/second versus third generation 
scalars and $a_3\equiv a_{Q_3H_uU_3}$ performs the same task for 
trilinear soft terms. The GMM$^\prime$ model
is programmed into the spectrum generator of Isajet\cite{isajet} which we
use for our sparticle mass calculations.
For simplicity, we take $c_m=c_{m3}=(5\ {\rm TeV}/\alpha M_s )^2$ so that 
matter scalar masses are $\sim 5$ TeV as in the NUHM2 case to be displayed
in Figs \ref{fig:mw1}{\it a}), \ref{fig:dew}{\it a}) and 
\ref{fig:lSUSY}{\it a}). We also take $a_3=1.6\sqrt{c_m}$.

One virtue of the LHC higgsino discovery plane is its relative model
independence.  Given some SUSY model, then for a given set of input
parameters one can calculate the (loop corrected)
values\cite{pp,gp,pbmz} of $m_{\tz_2}$ and $\Delta m^0$ and always
locate a point on the discovery plane.  Model dependence enters via the
assumed value of $m_{\tw_1}$.  The ATLAS and CMS groups assume
$m_{\twns_1^\pm}^{*}\equiv (m_{\tz_2}+m_{\tz_1})/2$ which roughly holds
at leading order in the deep higgsino region\cite{gp}.  When higher
order effects in the mass expansion or loop effects are included, then
there are deviations to this ansatz.  Since the details of the relative
chargino mass hardly affect the OSDJMET signature, the effects are not
so relevant, unless one begins leaving the nearly pure higgsino region
where $| \mu |\ll m_{soft}$.

As an illustration, we plot contours of mass difference $\Delta m(\tw_1
)\equiv m_{\twns_1^+}-m_{\twns_1^+}^*$ between the full one-loop
corrected chargino mass from Isajet and the ATLAS/CMS ansatz
$m_{\twns_1^+}^{*}$ in Fig. \ref{fig:mw1} for {\it a}) the NUHM2 model
and {\it b}) the GMM$^\prime$ model.  The blue contour has mass
difference zero so is an excellent fit to the ATLAS/CMS ansatz.
However, as one proceeds to higher $\mu$ values then the mass
differences becomes typically greater than zero with the chargino mass
becoming larger than the average of the two light neutralinos. For very
large $\mu$, then one leaves the light higgsino region and the ansatz no
longer obtains. 
The deviation of the chargino mass from the assumed simplified model value 
is not very relevant for the monojet plus soft dilepton searches 
considered below, but would be important for signals
such as the golden trilepton signal for SUSY
that originate from chargino-neutralino production\cite{Baer:2013xua}.

In Fig. \ref{fig:dew}, we show some aspects of the higgsino discovery
plane that are beyond the purview of the ATLAS/CMS simplified models and
which depend on the entire SUSY particle mass spectrum.  In
Fig. \ref{fig:dew}{\it a}), we scan over the NUHM2 parameters $\mu
:50-1000$ GeV (which fixes the higgsino masses) and $m_{1/2}: 100-2000$
GeV (which for a given $\mu$ value varies the mass gap $\Delta m^0$).
The remaining parameters are fixed at $m_0=5$ TeV, $A_0=-1.6 m_0$,
$\tan\beta =10$ and $m_A=2$ TeV.  Since the entire SUSY spectrum is
calculated, now we can compute the corresponding value of $\Delta_{EW}$
for each point in the higgsino discovery plane.  The green points have
$\Delta_{EW}<15$ while magenta points have $\Delta_{EW}<30$ and hence
qualify as natural.  Yellow, blue and purple points have
$\Delta_{EW}<100$, 200 and 300 respectively.  The grey-shaded region is
already excluded by LEP2 searches for chargino pair production.
From the plot, we see of course that the natural region is bounded by
$m_{\tz_2}\alt 350$ GeV as expected.  For small $m_{1/2}$ and $\mu >350$
GeV, then the $\tz_2$ is actually wino-like and the model can become
unnatural even for lower values of $m_{\tz_2}\sim 100-300$ GeV (which
forms the upper edge of the naturalness envelope in
Fig. \ref{fig:dew}{\it a})).  For fixed $\mu\sim 100-300$ GeV-- but as
$m_{1/2}$ increases-- then the lightest electroweakinos become
increasingly higgsino-like and the mass gap $\Delta m^0$ drops below
$\sim 8$ GeV.  The precise value of the mass gap where the model starts
to become unnatural is somewhat sensitive to the assumptions of the
NUHM2 model. Indeed somewhat lower values of the neutralino gap would
have $\Delta_{EW} \alt 30$ if we allow generation-dependent matter
scalar mass parameters, or if we give up the gaugino unification
assumption. The point, however, is that for small mass gaps, the points
become increasingly unnatural, in the NUHM2 case because large $m_{1/2}$
increases $m_{\tg}$ which feeds into the stop masses so that the
$\Sigma_u^u (\tst_{1,2})$ become too large.  Also, the two-loop
contributions from $m_{\tg}$ and $m_{\tst_{1,2}}$ can become
large\cite{Dedes:2002dy}.  This gives an important result: the region of
higgsino discovery plane with mass gaps $\Delta m^0\alt 5$ GeV becomes
increasingly unnatural and hence less plausible.  As mentioned above,
the naturalness lower bound on $\Delta m^0$ is somewhat model-dependent
and can reach as low as $\sim 4$ GeV in models like NUHM3 where
first/second generation matter sfermions take values in the 20-40 TeV
range. In that case, two-loop RGE effects suppress top squark soft term
running\cite{Baer:2000xa}, which allows larger $m_{1/2}$ values to be natural: these same
large $m_{1/2}$ values lead to smaller neutralino mass gaps $\Delta
m^0$.  While searches in this unnatural region of very low $\Delta m^0$
are always warranted, spending an inordinate effort probing tiny mass
gaps should be given a much lower priority in this rather implausible
region.\footnote{This is akin to the huge effort that went into placing
  limits on compressed stop-neutralino spectra in order to exclude
  natural SUSY, but under an overly-simplified measure of naturalness
  which emphasized (wrongly) that top squarks must be not too far
  removed from the weak scale.}

We also show in Fig. \ref{fig:dew}{\it a}) the corresponding contour of
$m_{\tg}=2.25$ TeV, the limit from ATLAS/CMS simplified model searches
for gluino pair production.  The region {\it above} the contour has
$m_{\tg}<2.25$ TeV and hence is largely excluded in the NUHM2 framework.
We emphasize that this exclusion directly depends on our assumption of
gaugino mass unification, and in more general models, the allowed
natural region may be considerably larger.  We also show the present
ATLAS search contour for the OSDJMET channel as the black contour.  The
region to the left of the contour is thus excluded.  Thus, the {\it
  allowed} NUHM2 natural search region has mass gap in the $\sim$7-20
GeV range, and this is the region where a SUSY signal may be expected.
The lower bound depends on the specific parameter choices adopted and
can range down to 4-5 GeV for other parameter choices.  The current
search results do cut well into the natural region of the NUHM2 model.
We also remind the reader that the ATLAS search has yielded a slight
excess in several invariant mass bins $m(\ell^+\ell^-)\sim 5-10$ GeV of
this search channel.  The projected reach of HL-LHC for CMS is shown by
the red contour, while the ATLAS HL-LHC projection is labelled by the
blue contour.  Some of the natural region of the higgsino discovery
plane lies beyond the HL-LHC projected reaches.  The ATLAS reach extends
to lower mass gaps evidently due to the geometry of the ATLAS detector
which allows for resolution of lower $p_T$ leptons than CMS.  The
projected reach of HE-LHC with $\sqrt{s}=27$ TeV for CMS is given by the
dashed red contour\cite{Canepa:2020ntc}.  The increased reach of HE-LHC
is mainly due to the assumed increase in potential integrated luminosity
when proceeding from HL- to HE-LHC: 3 ab$^{-1}$ $\to$ 15 ab$^{-1}$.  At
face value, the
projected HE-LHC reach apparently covers all the natural region of the
NUHM2 model for the assumed set of parameters.

In Fig. \ref{fig:dew}{\it b}), we show the same higgsino discovery plane
but for the GMM$^\prime$ model where the mirage-mediation (MM) value
$m_{1/2}^{MM}\equiv \alpha M_s$ again varies between 100-2000 GeV.  For
lower values of $m_{1/2}^{MM}$ we obtain tachyonic spectra (see Fig. 8
of Ref. \cite{Baer:2019tee}) so that no upper edge of unnaturalness
ensues as it did in Fig. \ref{fig:dew}{\it a} until $m_{\tz_2}\agt 400$
GeV.  For GMM$^\prime$, depending on $\alpha$, we may have a compressed
spectrum of gaugino masses as expected from mirage mediation. This means
that for a given value of $m_{\tg}$, the wino and bino masses can be
much larger than in the corresponding NUHM2 case with unified gaugino
masses. The large wino/bino masses in GMM$^\prime$ lead to smaller mass
gaps and in fact here we find natural spectra with mass gaps down to
$\Delta m^0\sim 6$ GeV.  In this case, more of the natural region is
explored by the ATLAS rather than CMS cuts and indeed more of the
natural SUSY parameter space appears to lie beyond HL-LHC reach.  Even a
tiny corner of magenta region seems to lie beyond projected HE-LHC
reach.  As in Fig. \ref{fig:dew}{\it a}), the region with mass gap
$\Delta m^0\alt 4-5$ GeV becomes increasingly unnatural.

From Fig. \ref{fig:dew}, it appears much if not most of the nature SUSY
parameter space is now excluded, including the values with lowest
$\Delta_{EW}$. This is a reflection of the $\Delta_{EW}$ measure which
is a bottom-up measure of {\it practical naturalness}: each of the independent
contributions $o_i$ to an observable {\cal O} ought to be comparable to
or less than its measured value. In contrast, from the successful
application of the statistics of string theory vacua to the prediction
of the cosmological constant (CC), the notion of {\it stringy
  naturalness} has arisen\cite{douglas,land4}: 
the value of an observable ${\cal O}_1$ is
more natural than the value ${\cal O}_2$ if more phenomenologically
viable vacua lead to ${\cal O}_1$ than to ${\cal O}_2$.  For the case of
the CC, for a uniform distribution of CC values $\Lambda$, then
statistical selection of pocket universes within the multiverse favor a
value of $\Lambda$ nearly as large as possible such that galaxies
condense, and structure forms in the universe. This reasoning allowed
Weinberg to predict the value of $\Lambda$ to within a factor of several
well before its value was measured\cite{Weinberg:1987dv}.

Applying similar reasoning to the SUSY breaking scale as expected from
string theory, then with a number of hidden sectors available, the
magnitude of the SUSY breaking scale is expected to scale as a {\it
  power law}\cite{Douglas:2004qg}: 
$dN_{vac}\sim m_{soft}^n$ where $n={2n_F+n_D-1}$.  This is
just a consequence of the fact that in string theory no particular SUSY
breaking vev is favored, so all values are equally likely. Then the
probability for the cumulative scale of SUSY breaking is just determined
by the dimensionality of the space of SUSY breaking fields, which
includes a factor of 2 for complex $F$-term breaking vevs and a factor 1
for real $D$-term breaking fields (as emphasized by Douglas and
others\cite{Douglas:2004qg,Susskind:2004uv,ArkaniHamed:2005yv}).
Already for SUSY breaking by a single $F$-term field, there is a linear
statistical draw towards large soft terms.  However, phenomenological
viability must also be addressed.  In the case of 4-d SUSY theories
containing the MSSM, the magnitude of the weak scale $m_{weak}$ is
determined by the values of soft breaking SUSY parameters and the
superpotential $\mu$ term.  Roughly, the larger the SUSY breaking scale,
then the larger is the associated value of the pocket-unverse weak scale
$m_{weak}^{PU}$. Agrawal {\it et al.}\cite{Agrawal:1997gf} 
have used nuclear physics calculations to argue that in order 
for complex nuclei to form, and hence atoms as we know them, 
then the PU value of the weak scale must be
within a factor 2-5 of the measured value of the weak scale in our
universe: $m_{weak}^{PU}\alt (2-5)m_Z^{OU}$.  For smoothly
distributed values of the $\mu$ term and SUSY breaking scale, this
amounts to $\Delta_{EW}<8-50$. For simplicity, we adopt an intermediate
value within this range: $\Delta_{EW}<30$ to yield a phenomenologically
viable weak scale.

In Fig. \ref{fig:lSUSY}, we adopt a value of $n=1$ for the gaugino
masses since in a wide variety of string models the gaugino masses
depend only on the dilaton field $S$ gaining a vev, whereas the various
moduli contribute subdominantly (the moduli and dilaton are expected to
contribute comparably to other soft terms such as trilinears and scalar
soft masses)\cite{Baer:2020vad}.  
We sample soft terms according to stringy naturalness with $n=1$ 
for gaugino masses but with a uniform distribution in $\mu$ (since the $\mu$
parameter arises from whatever solution to the SUSY $\mu$ problem is 
assumed\cite{Bae:2019dgg}) starting at $\mu >100$ GeV.
The resulting distribution of dots is displayed in Fig.~\ref{fig:lSUSY}. 
The density of dots is important in this case and higher density corresponds to
greater stringy naturalness.

In the case of the NUHM2 model displayed in Fig. \ref{fig:lSUSY}{\it
  a}), we see that the region of parameter space with small mass gap is
favored by stringy naturalness over the regions with large mass
gap. Thus, much of the stringy natural region still lies well beyond the
present reach of LHC. This is consistent with the statistical
predictions of stringy naturalness for the sparticle mass spectra:
stringy naturalness pulls the Higgs mass $m_h$ to a peak around 125 GeV
while gluinos are pulled up to $m_{\tg}\sim 4\pm 2$ TeV and stops to
$m_{\tst_1}\sim 1.5\pm 0.5$ TeV\cite{land1,land2,Baer:2019tee}.  Thus,
stringy naturalness seems to explain what LHC is seeing: a Higgs of mass
$m_h\simeq 125$ GeV with sparticles pulled beyond the present LHC reach.
For a fixed value of $\mu$, since stringy naturalness pulls the gaugino
masses as large as possible -- subject to maintaining $m_{weak}^{PU}$
not too far removed from our measured value $m_{weak}^{OU}$ -- then we
expect the $m_{\tz_2}-m_{\tz_1}$ mass gap to be favored on the low
allowed side: $\Delta m^0\sim 5-10$ GeV. A similar calculation performed
within the GMM$^{\prime}$ model yields similar results in
Fig. \ref{fig:lSUSY}{\it b}): the low mass gap region is statistically
favored within phenomenologically viable vacua within the multiverse.

Before concluding, it seems worthwhile to highlight the similarities and
differences between the naturalness considerations in Fig.~\ref{fig:dew}
and Fig.~\ref{fig:lSUSY}. The bottom-up measure $\Delta_{EW}$ is
universal and applies independently of the details of UV physics. In
contrast, the notion of stringy naturalness hinges on the existence of
string vacua and their distribution as well as on the atomic principle 
that led to the cut-off, $m_{weak}^{PU}\alt (2-5)m_Z^{OU}$. These
additional hypotheses about the nature of UV physics lead to a
preference for lower values of $\Delta m^0$. We stress, however, that
stringy naturalness together with the atomic principle is entirely 
compatible with electroweak naturalness. This is reflected in the fact
that the envelope of points in Fig.~\ref{fig:lSUSY} is essentially the 
same as that in Fig.~\ref{fig:dew}. The reader who does not subscribe to
the notion of stringy naturalness can simply disregard the preference
for points with lower $\Delta m^0$ apparent in
Fig.~\ref{fig:lSUSY}. 
However, the important conclusion that naturalness considerations 
require the neutralino mass gap to be not much below 4-5~GeV remains 
unaltered.\footnote{A previous paper explored the compressed electroweakino 
mass spectrum from natural SUSY with an eye towards the possibility of 
long-lived charginos with sub-GeV mass gaps\cite{Bomark:2013nya}. 
This work took place in the pMSSM11 model with upper limits on 
parameter choices arising from different notions of naturalness\cite{prw}. 
They also concluded that the mass gap should be larger than 5 GeV, and
emphasized that metastable higgsinos would {\it not} be a signature of 
(that version of) natural SUSY.
}

{\it Conclusions:} Based on electroweak naturalness and even more on
stringy naturalness, it may well be that gluinos and squarks, including
top squarks, lie well beyond the reach of HL-LHC, and so may have to
await an energy upgrade of LHC into the 27-50-100~TeV range for their
discovery.\footnote{It is worth noting that in natural SUSY models with
  $\Delta_{EW} < 30$, a 27~TeV $pp$ collider with an integrated
  luminosity of 15~ab$^{-1}$ would discover at least one of the stop or
  the gluino, and possibly both; discovery of other squarks and sleptons
  may have to await yet higher energy colliders.\cite{jamie}.}
In contrast, the lightest electroweakinos are expected to be
mainly higgsino-like with masses not too far removed from the measured
value of the weak scale $m_{weak}\simeq m_{W,Z,h}\sim 100$ GeV. Thus,
higgsino pair production is expected to occur at considerable rates at
HL-LHC.  The problem instead is one of visible energy: the small mass
gaps $m_{\tw_1}-m_{\tz_1}$ and especially $m_{\tz_2}-m_{\tz_1}$ are
expected to be on the $5-10$ GeV range and most of the reaction energy
goes into making the LSP masses $2m_{\tz_1}$.  In such a case, it
appears the soft opposite-sign dilepton plus jet plus MET signature
OSDJMET is most promising, which depends on initial state radiation of a
hard gluon or quark jet so that MET or $p_T(jet)$ can be used as a
trigger.  ATLAS and CMS have been analyzing these reactions and plotting
excluded regions in the simplified model $m_{\tz_2}$ vs. $\Delta m^0$
plane and in fact ATLAS has a slight excess of events in this channel
with $m(\ell^+\ell^- )\sim 4-12$ GeV from 139 fb$^{-1}$ of data.  From
the theory perspective, not all parts of the higgsino discovery plane
are equally plausible. In this paper we plotted out the natural portions
of the discovery plane using the model-independent naturalness measure
$\Delta_{EW}$.  Large portions of the natural region is already excluded
by both gluino pair searches and by the OSDJMET search channel. However,
considerable portions of the discovery plane remain unconstrained,
especially those with low mass gaps $\Delta m^0\sim 5-10$ GeV.  Indeed
these very portions are most favored by stringy naturalness, which also
predicts $m_h\sim 125$ GeV with sparticles beyond the present LHC reach
(along with the magnitude of the CC). Thus, experimental searches may
wish to concentrate on these (stringy) natural regions, with perhaps
lower priority efforts directed to mass gaps significantly below 4-5~GeV
and certainly below 1 GeV. In those regions, huge gaugino masses are
required which ultimately spoil the naturalness of the models.

{\it Acknowledgements:} 
We thank A. Canepa and  T. Han for discussions.
This work was supported in part by the US Department of Energy, Office
of High Energy Physics. 


%

\begin{thebibliography}{99}


\bibitem{atlas_h} G.~Aad {\it et al.}  [ATLAS Collaboration], 
 Phys. Lett. {\bf B716} (2012) 1.
%
\bibitem{cms_h} S.~Chatrchyan {\it et al.}  [CMS Collaboration],  
Phys. Lett. {\bf B716} (2012) 30.
%
\bibitem{witten_kaul} 
  E.~Witten,
  Nucl.\ Phys.\ B {\bf 188} (1981) 513;
R.~K.~Kaul,
  Phys.\ Lett.\  {\bf 109B} (1982) 19.

\bibitem{wss} H.~Baer and X.~Tata,
  ``Weak scale supersymmetry: From superfields to scattering events,''
  Cambridge, UK: Univ. Pr. (2006) 537 p.
%

\bibitem{Broeckel:2020fdz}
I.~Broeckel, M.~Cicoli, A.~Maharana, K.~Singh and K.~Sinha,
[arXiv:2007.04327 [hep-th]].


%
\bibitem{gauge} S.~Dimopoulos, S.~Raby and F.~Wilczek,
  Phys.\ Rev.\ D {\bf 24} (1981) 1681;
K.~Inoue, A.~Kakuto, H.~Komatsu and S.~Takeshita,
  Prog.\ Theor.\ Phys.\  {\bf 68} (1982) 927
   [Erratum-ibid.\  {\bf 70} (1983) 330]
   [Prog.\ Theor.\ Phys.\  {\bf 70} (1983) 330];
L.~Alvarez-Gaume, J.~Polchinski and M.~B.~Wise,
  Nucl.\ Phys.\ B {\bf 221} (1983) 495.
K.~Inoue, A.~Kakuto, H.~Komatsu and S.~Takeshita,
  Prog.\ Theor.\ Phys.\  {\bf 71} (1984) 413.
%
\bibitem{rewsb} L. E. Iba\~nez and G. G. Ross, Phys. Lett. {\bf B110}, 215
(1982); K. Inoue {\it et al.} Prog. Theor. Phys. {\bf 68}, 927 (1982)
and {\bf 71}, 413 (1984); 
L.~Iba\~nez, Phys. Lett. {\bf B118}, 73 (1982); 
 H.~P.~Nilles, M.~Srednicki and D.~Wyler,
  Phys.\ Lett.\ B {\bf 120} (1983) 346;
J.~Ellis, J.~Hagelin, D.~Nanopoulos and M.~Tamvakis,
Phys. Lett. {\bf B125}, 275 (1983); 
L.~Alvarez-Gaum\'e. J.~Polchinski and M.~Wise,
Nucl. Phys. {\bf B221}, 495 (1983);
B.~A.~Ovrut and S.~Raby,
  Phys.\ Lett.\ B {\bf 130} (1983) 277;
for a review, see 
L.~E.~Ibanez and G.~G.~Ross,
  Comptes Rendus Physique {\bf 8} (2007) 1013.
%
\bibitem{mhiggs} M.~Carena and H.~E.~Haber,
``Higgs boson theory and phenomenology,''
  Prog.\ Part.\ Nucl.\ Phys.\  {\bf 50}, 63 (2003);
P.~Draper and H.~Rzehak,
``A Review of Higgs Mass Calculations in Supersymmetric Models,''
  Phys.\ Rept.\  {\bf 619}, 1 (2016).
%
\bibitem{sven} S.~Heinemeyer, W.~Hollik, D.~Stockinger, A.~M.~Weber and G.~Weiglein,
  JHEP {\bf 0608} (2006) 052
  doi:10.1088/1126-6708/2006/08/052
  [hep-ph/0604147].
%
\bibitem{Canepa:2019hph}
  A.~Canepa,
  Rev.\ Phys.\  {\bf 4} (2019) 100033.
  doi:10.1016/j.revip.2019.100033
%
\bibitem{lhc_gl} M.~Aaboud {\it et al.} [ATLAS Collaboration],
  Phys.\ Rev.\ D {\bf 97} (2018) no.11,  112001
  doi:10.1103/PhysRevD.97.112001
  [arXiv:1712.02332 [hep-ex]]; 
T.~A.~Vami [ATLAS and CMS Collaborations],
  PoS LHCP {\bf 2019} (2019) 168
  doi:10.22323/1.350.0168
  [arXiv:1909.11753 [hep-ex]].
%
\bibitem{lhc_t1} The ATLAS collaboration [ATLAS Collaboration],
  ATLAS-CONF-2019-017; 
A.~M.~Sirunyan {\it et al.} [CMS Collaboration],
  arXiv:1912.08887 [hep-ex].
%
\bibitem{eenz} J.~R.~Ellis, K.~Enqvist, D.~V.~Nanopoulos and F.~Zwirner,
Mod.\ Phys.\ Lett.\ A {\bf 1}, 57 (1986).
%
\bibitem{bg} R.~Barbieri and G.~F.~Giudice,
Nucl.\ Phys.\ B {\bf 306}, 63 (1988).
%
\bibitem{dg} S.~Dimopoulos and G.~F.~Giudice,
  Phys.\ Lett.\ B {\bf 357} (1995) 573.
%
\bibitem{ac} G.~W.~Anderson and D.~J.~Castano,
Phys.\ Rev.\ D {\bf 53} (1996) 2403.
%
\bibitem{DEW} H.~Baer, V.~Barger and D.~Mickelson,
Phys.\ Rev.\ D {\bf 88}, 095013 (2013).
%
\bibitem{mt} A.~Mustafayev and X.~Tata,
  Indian J.\ Phys.\  {\bf 88} (2014) 991.
%
\bibitem{seige}
  H.~Baer, V.~Barger, D.~Mickelson and M.~Padeffke-Kirkland,
  Phys.\ Rev.\ D {\bf 89} (2014) no.11,  115019
  doi:10.1103/PhysRevD.89.115019
  [arXiv:1404.2277 [hep-ph]].
%
\bibitem{arno} H.~Baer, V.~Barger and M.~Savoy,
Phys. Scripta \textbf{90} (2015), 068003
doi:10.1088/0031-8949/90/6/068003
[arXiv:1502.04127 [hep-ph]].
%
\bibitem{midi} H.~Baer, V.~Barger, S.~Salam, D.~Sengupta and K.~Sinha,
[arXiv:2002.03013 [hep-ph]].
%
\bibitem{land1} H.~Baer, V.~Barger, M.~Savoy and H.~Serce,
Phys. Lett. B \textbf{758} (2016), 113-117
doi:10.1016/j.physletb.2016.05.010
[arXiv:1602.07697 [hep-ph]].
%
\bibitem{land2} H.~Baer, V.~Barger, H.~Serce and K.~Sinha,
JHEP \textbf{03} (2018), 002
doi:10.1007/JHEP03(2018)002
[arXiv:1712.01399 [hep-ph]].
%
\bibitem{land3} H.~Baer, V.~Barger, S.~Salam, H.~Serce and K.~Sinha,
JHEP \textbf{04} (2019), 043
doi:10.1007/JHEP04(2019)043
[arXiv:1901.11060 [hep-ph]].
%
\bibitem{land4} H.~Baer, V.~Barger and S.~Salam,
Phys. Rev. Research. \textbf{1} (2019), 023001
doi:10.1103/PhysRevResearch.1.023001
[arXiv:1906.07741 [hep-ph]].
%
\bibitem{ltr} H.~Baer, V.~Barger, P.~Huang, A.~Mustafayev and X.~Tata,
Phys. Rev. Lett. {\bf 109}, 161802 (2012).
%
\bibitem{rns} H.~Baer, V.~Barger, P.~Huang, D.~Mickelson, A.~Mustafayev and X.~Tata,
Phys.\ Rev.\ D {\bf 87}, 115028 (2013).
%
 \bibitem{isajet} F.~E.~Paige, S.~D.~Protopopescu, H.~Baer and X.~Tata,
  hep-ph/0312045.
%
\bibitem{Dedes:2002dy}
A.~Dedes and P.~Slavich,
Nucl. Phys. B \textbf{657} (2003), 333-354
doi:10.1016/S0550-3213(03)00173-1
[arXiv:hep-ph/0212132 [hep-ph]].
%
\bibitem{upper} H.~Baer, V.~Barger and M.~Savoy,
Phys. Rev. D \textbf{93} (2016) no.3, 035016
doi:10.1103/PhysRevD.93.035016
[arXiv:1509.02929 [hep-ph]].
%
\bibitem{gainer27} H.~Baer, V.~Barger, J.~S.~Gainer, D.~Sengupta, H.~Serce and X.~Tata,
Phys. Rev. D \textbf{98} (2018) no.7, 075010
doi:10.1103/PhysRevD.98.075010
[arXiv:1808.04844 [hep-ph]].
%
\bibitem{Douglas:2004qg}
  M.~R.~Douglas,
  hep-th/0405279.
%
\bibitem{Donoghue:2007zz}
J.~F.~Donoghue,
[arXiv:0710.4080 [hep-ph]].
%
\bibitem{Bae:2019dgg}
  K.~J.~Bae, H.~Baer, V.~Barger and D.~Sengupta,
  Phys.\ Rev.\ D {\bf 99} (2019) no.11,  115027
  doi:10.1103/PhysRevD.99.115027
  [arXiv:1902.10748 [hep-ph]].
%
\bibitem{douglas} M.~R.~Douglas,
  Comptes Rendus Physique {\bf 5} (2004) 965
  doi:10.1016/j.crhy.2004.09.008
  [hep-th/0409207].
%
\bibitem{Baer:2019zfl}
H.~Baer, V.~Barger and D.~Sengupta,
Phys. Rev. Res. \textbf{1} (2019) no.3, 033179
doi:10.1103/PhysRevResearch.1.033179
[arXiv:1910.00090 [hep-ph]].
%
\bibitem{Baer:2013vpa} H.~Baer, V.~Barger and D.~Mickelson,
  Phys.\ Lett.\ B {\bf 726} (2013) 330.
%
\bibitem{Baer:2017cck}
  H.~Baer, V.~Barger, M.~Savoy, H.~Serce and X.~Tata,
  JHEP {\bf 1706} (2017) 101
  doi:10.1007/JHEP06(2017)101
  [arXiv:1705.01578 [hep-ph]].
%
\bibitem{Bae:2013bva}
K.~J.~Bae, H.~Baer and E.~J.~Chun,
Phys. Rev. D \textbf{89} (2014) no.3, 031701
doi:10.1103/PhysRevD.89.031701
[arXiv:1309.0519 [hep-ph]].
%
\bibitem{cck} K.~Choi, E.~J.~Chun and J.~E.~Kim,
Phys. Lett. B \textbf{403} (1997), 209-217
doi:10.1016/S0370-2693(97)00465-6
[arXiv:hep-ph/9608222 [hep-ph]].
%
\bibitem{spm} S.~P.~Martin,
  Phys.\ Rev.\ D {\bf 54} (1996) 2340;
S.~P.~Martin,
  Phys.\ Rev.\ D {\bf 61} (2000) 035004; 
S.~P.~Martin,
  Phys.\ Rev.\ D {\bf 62} (2000) 095008.
%
\bibitem{Nilles:2017heg}
H.~P.~Nilles,
PoS \textbf{CORFU2016} (2017), 017
doi:10.22323/1.292.0017
[arXiv:1705.01798 [hep-ph]].
%
\bibitem{Baer:2018avn}
H.~Baer, V.~Barger and D.~Sengupta,
Phys. Lett. B \textbf{790} (2019), 58-63
doi:10.1016/j.physletb.2019.01.007
[arXiv:1810.03713 [hep-ph]].
%
\bibitem{gp} G.~F.~Giudice and A.~Pomarol,
Phys. Lett. B \textbf{372} (1996), 253-258
doi:10.1016/0370-2693(96)00060-3
[arXiv:hep-ph/9512337 [hep-ph]].
%
\bibitem{Baer:2014cua}
H.~Baer, A.~Mustafayev and X.~Tata,
Phys. Rev. D \textbf{89} (2014) no.5, 055007
doi:10.1103/PhysRevD.89.055007
[arXiv:1401.1162 [hep-ph]]; C.~Han, A.~Kobakhidze, N.~Liu, A.~Saavedra,
L.~Wu and J.~M.~Yang, JHEP {\bf 1402} (2014) 049
[arXiv:1310.4274[hep-ph]]; P.~Schwaller and J.~Zurita, JHEP {\bf 1403}
(2014) 060 [arXiv:1312.7350 [hep-ph]].

%
\bibitem{bbh} H.~Baer, V.~Barger and P.~Huang,
  JHEP {\bf 1111} (2011) 031.
%
\bibitem{than} G.~F.~Giudice, T.~Han, K.~Wang and L.~T.~Wang,
Phys. Rev. D \textbf{81} (2010), 115011
doi:10.1103/PhysRevD.81.115011
[arXiv:1004.4902 [hep-ph]].
%
\bibitem{kribs} Z.~Han, G.~D.~Kribs, A.~Martin and A.~Menon,
``Hunting quasidegenerate Higgsinos,'' 
  Phys.\ Rev.\ D {\bf 89} no.7, 075007 (2014);
H.~Baer, A.~Mustafayev and X.~Tata,
``Monojet plus soft dilepton signal from light higgsino pair production at LHC14,''                
  Phys.\ Rev.\ D {\bf 90} no.11, 115007 (2014);
C.~Han, D.~Kim, S.~Munir and M.~Park,
``Accessing the core of naturalness, nearly degenerate higgsinos, at the LHC,'' JHEP {\bf 1504}, 132 (2015);
H.~Baer, V.~Barger, M.~Savoy and X.~Tata,
``Multichannel assault on natural supersymmetry at the high luminosity LHC,''
  Phys.\ Rev.\ D {\bf 94} no.3, 035025 (2016).
%
\bibitem{Aad:2019qnd}
G.~Aad \textit{et al.} [ATLAS],
Phys. Rev. D \textbf{101} (2020) no.5, 052005
doi:10.1103/PhysRevD.101.052005
[arXiv:1911.12606 [hep-ex]].
%
\bibitem{CMS:2016zvj}
CMS Collaboration,
CMS-PAS-SUS-16-025.
%
\bibitem{nuhm2}  D.~Matalliotakis and H.~P.~Nilles,
  Nucl.\ Phys.\ B {\bf 435} (1995) 115;
M.~Olechowski and S.~Pokorski,
  Phys.\ Lett.\ B {\bf 344} (1995) 201;
P.~Nath and R.~L.~Arnowitt,
  Phys.\ Rev.\ D {\bf 56} (1997) 2820;
J. Ellis, K. Olive and Y. Santoso, Phys. Lett. {\bf B539} (2002) 107;
J. Ellis, T. Falk, K. Olive and Y. Santoso, 
Nucl. Phys. {\bf B652} (2003) 259;
H.~Baer, A.~Mustafayev, S.~Profumo, A.~Belyaev and X. Tata, 
JHEP{\bf 0507} (2005) 065.
%
\bibitem{raby} S.~Raby,
Rept. Prog. Phys. \textbf{74} (2011), 036901
doi:10.1088/0034-4885/74/3/036901
[arXiv:1101.2457 [hep-ph]].
%
\bibitem{localguts} W.~Buchmuller, K.~Hamaguchi, O.~Lebedev and M.~Ratz,
[arXiv:hep-ph/0512326 [hep-ph]];
M.~Ratz,
Soryushiron Kenkyu Electron. \textbf{116} (2008), A56-A76
doi:10.24532/soken.116.1-A56
[arXiv:0711.1582 [hep-ph]].
%
\bibitem{GMM}
  H.~Baer, V.~Barger, H.~Serce and X.~Tata,
  Phys.\ Rev.\ D {\bf 94} (2016) no.11,  115017
  doi:10.1103/PhysRevD.94.115017
  [arXiv:1610.06205 [hep-ph]].
%
\bibitem{amsb} L.~Randall and R.~Sundrum,
  Nucl.\ Phys.\ B {\bf 557} (1999) 79
  doi:10.1016/S0550-3213(99)00359-4
  [hep-th/9810155];
G.~F.~Giudice, M.~A.~Luty, H.~Murayama and R.~Rattazzi,
  JHEP {\bf 9812} (1998) 027
  doi:10.1088/1126-6708/1998/12/027
  [hep-ph/9810442];
J.~A.~Bagger, T.~Moroi and E.~Poppitz,
  JHEP {\bf 0004} (2000) 009
  doi:10.1088/1126-6708/2000/04/009
  [hep-th/9911029].
%
\bibitem{Choi:2005uz}
K.~Choi, K.~S.~Jeong and K.~i.~Okumura,
JHEP \textbf{09} (2005), 039
doi:10.1088/1126-6708/2005/09/039
[arXiv:hep-ph/0504037 [hep-ph]].
%
\bibitem{Falkowski:2005ck}
  A.~Falkowski, O.~Lebedev and Y.~Mambrini,
  JHEP {\bf 0511} (2005) 034
  doi:10.1088/1126-6708/2005/11/034
  [hep-ph/0507110].
%
\bibitem{Choi:2007ka}
  K.~Choi and H.~P.~Nilles,
  JHEP {\bf 0704} (2007) 006
  doi:10.1088/1126-6708/2007/04/006
  [hep-ph/0702146 [HEP-PH]].
%
\bibitem{Choi:2006xb}
  K.~Choi, K.~S.~Jeong, T.~Kobayashi and K.~i.~Okumura,
  Phys.\ Rev.\ D {\bf 75} (2007) 095012
  doi:10.1103/PhysRevD.75.095012
  [hep-ph/0612258].
%
\bibitem{pp} D.~Pierce and A.~Papadopoulos,
Nucl. Phys. B \textbf{430} (1994), 278-294
doi:10.1016/0550-3213(94)00303-3
[arXiv:hep-ph/9403240 [hep-ph]].
%
\bibitem{pbmz} D.~M.~Pierce, J.~A.~Bagger, K.~T.~Matchev and R.~j.~Zhang,
Nucl. Phys. B \textbf{491} (1997), 3-67
doi:10.1016/S0550-3213(96)00683-9
[arXiv:hep-ph/9606211 [hep-ph]].
%
\bibitem{Baer:2013xua}
H.~Baer, V.~Barger, P.~Huang, D.~Mickelson, A.~Mustafayev, W.~Sreethawong and X.~Tata,
JHEP \textbf{12} (2013), 013
doi:10.1007/JHEP12(2013)013
[arXiv:1310.4858 [hep-ph]].
%
\bibitem{Baer:2000xa}
H.~Baer, C.~Balazs, P.~Mercadante, X.~Tata and Y.~Wang,
Phys. Rev. D \textbf{63} (2001), 015011
doi:10.1103/PhysRevD.63.015011
[arXiv:hep-ph/0008061 [hep-ph]].
%
\bibitem{Canepa:2020ntc}
A.~Canepa, T.~Han and X.~Wang,
doi:10.1146/annurev-nucl-031020-121031
[arXiv:2003.05450 [hep-ph]].
%
\bibitem{Baer:2019tee}
H.~Baer, V.~Barger and D.~Sengupta,
Phys. Rev. Res. \textbf{2} (2020) no.1, 013346
doi:10.1103/PhysRevResearch.2.013346
[arXiv:1912.01672 [hep-ph]].
%
\bibitem{Weinberg:1987dv}
  S.~Weinberg,
  Phys.\ Rev.\ Lett.\  {\bf 59} (1987) 2607.
  doi:10.1103/PhysRevLett.59.2607;
  S.~Weinberg,
  Rev.\ Mod.\ Phys.\  {\bf 61} (1989) 1.
  doi:10.1103/RevModPhys.61.1
%
\bibitem{Susskind:2004uv}
  L.~Susskind,
  In *Shifman, M. (ed.) et al.: From fields to strings, vol. 3* 1745-1749
  doi:10.1142/9789812775344-0040
  [hep-th/0405189].
%
\bibitem{ArkaniHamed:2005yv}
N.~Arkani-Hamed, S.~Dimopoulos and S.~Kachru,
[arXiv:hep-th/0501082 [hep-th]].
%
\bibitem{Agrawal:1997gf}
  V.~Agrawal, S.~M.~Barr, J.~F.~Donoghue and D.~Seckel,
  Phys.\ Rev.\ Lett.\  {\bf 80} (1998) 1822
  doi:10.1103/PhysRevLett.80.1822
  [hep-ph/9801253];
V.~Agrawal, S.~M.~Barr, J.~F.~Donoghue and D.~Seckel,
  Phys.\ Rev.\ D {\bf 57} (1998) 5480
  doi:10.1103/PhysRevD.57.5480
  [hep-ph/9707380].
%
\bibitem{Baer:2020vad}
H.~Baer, V.~Barger, S.~Salam and D.~Sengupta,
[arXiv:2005.13577 [hep-ph]].
%
\bibitem{jamie} H.~Baer, V.~Barger, J.~Gainer and D.~Sengupta, H.~Serce
  and X.~Tata, Phys.\ Rev.\ D {\bf 98} (2018) 7, 075010
  [arXiv:1808.04844 [hep-ph]].
%
\bibitem{Bomark:2013nya}
N.~E.~Bomark, A.~Kvellestad, S.~Lola, P.~Osland and A.~R.~Raklev,
JHEP \textbf{05} (2014), 007
doi:10.1007/JHEP05(2014)007
[arXiv:1310.2788 [hep-ph]].
%
\bibitem{prw} M.~Papucci, J.~T.~Ruderman and A.~Weiler,
  JHEP {\bf 1209} (2012) 035.
%
\end{thebibliography}
\end{document}